\documentclass[twoside]{article}
\usepackage{fleqn,espcrc2}

\usepackage{graphicx}

\def\fkm {P. Fisher, B. Kayser, and K.S. McFarland, {\it Ann.\ Rev.\ Nucl.\ Part.\ Sci.} {\bf 49}, eds. C. Quigg, V. Luth, and P. Paul (Annual Reviews, Palo Alto, CA, 1999) 481}
\def\plb#1#2#3{{\it Phys.\ Lett.} {\bf B#1} (19#2) #3}
\def\prdn#1#2#3{{\it Phys.\ Rev.} {\bf D#1} (20#2) #3}
\def\prd#1#2#3{{\it Phys.\ Rev.} {\bf D#1} (19#2) #3}
\def\prl#1#2#3{{\it Phys.\ Rev.\ Lett.} {\bf #1} (19#2) #3}

\def\st{\scriptstyle}
\def\ss{\scriptsize}
\def\sss{\scriptscriptstyle}
\def\barp{{\raise.35ex\hbox{${\sss (}$}}---{\raise.35ex\hbox{${\sss )}$}}}
\def\nlbarp{\hbox{$\nu_\ell$\kern-1.0em\raise1.4ex\hbox{\barp}}}
\def\nlpbarp{\hbox{$\nu_{\ell^{\prime}}$\kern-1.4em \raise1.4ex\hbox{\barp}}}
\def\barq{{\raise.30ex\hbox{${\sss ``}$}}---{\raise.30ex\hbox{${\sss "}$}}}
\def\nbarq{\hbox{$\nu_m$\kern-1.4em\raise1.4ex\hbox{\barq}}}
\def\nmb{\overline{\nu_m}}

\def\dm2{\delta M^2_{ m \, m^{\sss\prime}}}
\def\DCP#1{\Delta_{CP} (#1)}

\def\ket#1{|#1\rangle}

\def\ra{\rightarrow}
\def\decayarrow{\kern0.2em\hbox{$\raise1.08ex\hbox{\big|}\kern-0.5em
                \longrightarrow$}}

\def\gsim{\;\raisebox{-.6ex}{$\stackrel{>}{\sim}$}\;}
\def\pom{\raisebox{-.8ex}{$\stackrel{+}{{\sss (}-{\sss )}}$}}
\def\thingie{\hbox{\kern-9pt\raise1pt%
         \hbox{{\fiverm(}{\lower1.5pt\hbox{\twelvebf--}}{\fiverm)}}}}

\newcommand{\Mbb}{|M_{\beta\beta}|}
\newcommand{\bb}{\beta\beta_{0 \nu}}
\newcommand{\Eq}[1]{Eq.~(\ref{eq#1})}
\newcommand{\beq}{\begin{equation}}
\newcommand{\eeq}{\end{equation}}

\title{Neutrino Properties}

\author{B. Kayser\address{Division of Physics, National Science
Foundation \\ 4201 Wilson Blvd., Arlington VA 22230, USA}%
\thanks{To appear in the Proceedings of the XIX International Conference on Neutrino Physics and Astrophysics, held in Sudbury, Canada, June 2000.}}

\begin{document}

\begin{abstract}
The experimental discovery that neutrinos almost certainly have masses and mix raises a number of fundamental questions about the neutrinos. We discuss what is presently known about the answers to these questions, and how we can learn more.
\vspace{1pc}
\end{abstract}

\maketitle

%\section{PUT SECTION TITLE HERE}

Thanks to very strong evidence that neutrinos oscillate \cite{r1}, we now know that they almost certainly have masses and mix. That neutrinos have masses means that there is some spectrum of three or more neutrino mass eigenstates, $\nu_1, \nu_2, \nu_3, \ldots$, which are the neutrino analogues of the charged-lepton mass eigenstates, $e, \mu$, and $\tau$. That neutrinos mix means that the weak interaction couples a given charged-lepton mass eigenstate $\ell$ ($= e, \mu$, or $\tau$) to more than one neutrino mass eigenstate $\nu_m$.
For example, in the weak process $W^+ \ra \ell^+ \nu_m$ with a given $\ell$, the produced neutrino can be any of the mass eigenstates. The amplitude for it to be the specific mass eigenstate $\nu_m$, $U^*_{\ell m}$, is an element of the unitary leptonic mixing matrix, often referred to as the Maki-Nakagawa-Sakata \cite{r2} matrix. Since $W^+ \ra \ell^+ \nu_m$ with given $\ell$ yields the neutrino mass eigenstate $\nu_m$ with amplitude $U^*_{\ell m}$, the neutrino state $\ket{\nu_\ell}$ created in this process is
\beq
\ket{\nu_\ell} = \sum_m U^*_{\ell m} \ket{\nu_m} \,\,.
\label{eq1}
\eeq
This state, produced in association with the charged lepton of ``flavor'' $\ell$, is called the neutrino of flavor $\ell$.

Although there are only three charged leptons of definite mass, there may be more than three neutrinos of definite mass. If there are, say, four such neutrinos $\nu_m$, then one linear combination of them,
\beq
\ket{\nu_{\mbox{{\ss Sterile}}}} = \sum_m U^*_{s m} \ket{\nu_m} \; ,
\label{eq2}
\eeq
has no normal weak couplings. In consequence, this linear combination is referred to as a ``sterile'' neutrino. 

Having discovered that neutrinos have masses and mix, we would like to learn the answers to the following questions:
\begin{itemize}
\item How many neutrino flavors, active and sterile, are there? Equivalently, how many neutrino mass eigenstates are there?
\item What are the masses, $M_m$, of the mass eigenstates $\nu_m$?
\item Is each neutrino of definite mass a Majorana particle ($\nmb = \nu_m$), or a Dirac particle  ($\nmb \neq \nu_m$)?
\item What are the elements $U_{\ell m}$ of the leptonic mixing matrix?
\item Does the behavior of neutrinos, in oscillation and other contexts, violate CP invariance?
\item What are the electromagnetic properties of neutrinos? In particular, what are their dipole moments?
\item What are the lifetimes of the neutrinos?
\end{itemize}

Let us discuss each of these questions in turn, recalling what is already known about the question, and considering how we can learn more.

It is generally believed that if the solar, atmospheric, and LSND neutrinos all genuinely oscillate, then there must be more than three neutrinos. A number of attempts have been made to accommodate all three of these oscillations with just three neutrinos \cite{r3}. However, it has also been argued that such attempts cannot succeed \cite{r4}.
The basic reason is that each of the three reported oscillations requires a neutrino squared-mass splitting $\dm2 \equiv M^2_m - M^2_{m^{\st \prime}}$ which is of a different order of magnitude than the splittings required by the other two. This is summarized in the table below.

\vspace{.15in}
\begin{tabular}{@{}cc}   \hline
Oscillating Neutrinos		& Required $| \delta M^2 |$ (eV$^2$)  \\ \hline 
Solar						& Bet. $10^{-11}$ and $10^{-4}$  \\
Atmospheric			& Few $\times 10^{-3}$  \\
LSND						& $\sim 1$  \\  \hline
\end{tabular} \\ [8pt]
Now, if there are only three neutrino mass eigenstates, then there are only three distinct splittings $\dm2$, and they are related to each other by
\begin{eqnarray}
\lefteqn{\sum \dm2 =  } \nonumber  \\
 & & (M^2_3-M^2_2) + (M^2_2-M^2_1) + (M^2_1-M^2_3)  \nonumber \\
 & &  = 0 \; .
\label{eq3}
\end{eqnarray}
But if the $\delta M^2$ values called for by the three oscillations are of three different orders of magnitude, then, regardless of their signs, they cannot possibly satisfy this relation. Thus, to explain the solar, atmospheric, and LSND oscillations, we must add a fourth neutrino mass eigenstate. Now, the decays $Z\ra \nu_\ell \, \overline{\nu_\ell}$ yield only three distinct neutrinos of definite flavor. Thus, the four neutrino flavor eigenstates corresponding to the four mass eigenstates must include one which does not couple to the $Z$ and hence does not enjoy normal weak interactions. That is, the four flavor eigenstates muist be $\nu_e, \nu_\mu, \nu_\tau$, and a sterile neutrino $\nu_{\mbox{{\ss Sterile}}}$. Thus, if the solar, atmospheric, and LSND oscillations are all real, nature contains a fourth neutrino quite unlike the three neutrinos with which we are already familiar.

To determine the masses $M_m$ of the neutrinos of definite mass, we will have to appeal to experiments other than neutrino oscillation studies, because neutrino oscillation can determine only mass splittings. To see why, recall that the amplitude $A\,(\nu_\ell \ra \nu_{\ell ^{\sss \prime}})$ for a neutrino of flavor $\ell$ and energy $E$ to oscillate into one of flavor $\ell^\prime$ in a distance $L$ is given by
\beq
A\,(\nu_\ell \ra \nu_{\ell ^{\sss\prime}}) = \sum_m U^*_{\ell m} U_{\ell^{\sss\prime} m} e^{-iM^2_m \frac{L}{2E}} \; .
\label{eq4}
\eeq
Notice that in this amplitude, the neutrino masses $M_m$ occur only in phase factors. When  we calculate the oscillation probability $P\,(\nu_\ell \ra \nu_{\ell ^{\sss\prime}}) = |A\,(\nu_\ell \ra \nu_{\ell^ {\sss\prime}})|^2$, only {\em relative} phases will matter, so $P\,(\nu_\ell \ra \nu_{\ell ^{\sss\prime}})$ will depend only on the mass splittings $\dm2 \equiv M^2_m - M^2_{m^{\sss\prime}}$, and not on the individual masses \cite{r4a}.

An experiment which, in principle, can provide information on individual neutrino masses is the study of the electron energy spectrum in tritium $\beta$ decay, $^3$H $\ra \,^3$He$ \;+ \;e^-  + \overline{\nu}$. However, this study may not be able to achieve sensitivity to masses $M_m$ much below 1 eV \cite{r5}. Now, there may indeed be neutrino mass eigenstates with masses this large. In fact, if the LSND oscillation is genuine, then there must be at least one mass eigenstate, $\nu_H$, with mass $M_H \geq (\delta M^2_{\mbox{\ss LSND}})^{1/2} \gsim (0.2$ eV$^2)^{1/2} \simeq 0.4$ eV. Here, $\delta M^2_{\mbox{\ss LSND}}$ is the squared-mass splitting required to fit the LSND data \cite{r6}. 
Perhaps the mass $M_H \geq 0.4$ eV would be within range of improved tritium experiments of the future. However, the branching ratio for emission of $\nu_H$ in tritium decay, BR$(^3$H $\ra \;^3$He$ \;+\; e^-  + \overline{\nu_H}) \sim |U_{eH}|^2$, may or may not be large \cite{r7}. 
Moreover, should it turn out that the LSND oscillation is not genuine, then the heaviest mass eigenstate may have a mass no larger than $(\delta M^2_{\mbox{\ss Atmos}})^{1/2} \sim (3.2 \times 10^{-3} \mbox{ eV}^2)^{1/2} \sim 0.06$ eV. The latter mass, which appears to be out of reach of tritium experiments, is the minimum mass implied by the squared-mass splitting $\delta M^2_{\mbox{\ss Atmos}}$ required \cite{r8} to fit the atmospheric neutrino oscillation data. If the LSND oscillation is not real, then $\delta M^2_{\mbox{\ss Atmos}}$ is the largest splitting implied by the oscillation data.

The conclusion is that it is important to pursue the tritium experiments, since there may well be a neutrino with a mass in the 1 eV range, but it is also important to try to think of some clever approach that will make possible the measurement of individual neutrino masses far below 1 eV.

Is each neutrino of definite mass $\nu_m$ identical to its antiparticle? First of all, what does this question mean? We know that in the weak decay $W^+ \ra \ell^+ \nu_m(h)$, the outgoing neutral particle is emitted with left-handed helicity $h$: $h = -1/2$. By contrast, in the decay $W^- \ra \ell^- \nbarq (h)$, the neutral particle is emitted with right-handed helicity: $h = +1/2$. In the $W^+$ decay, the outgoing neutral particle is called a neutrino because it is produced together with a charged lepton of positive charge. 
In the $W^-$ decay, it is called an ``antineutrino'' because it is produced with a charged lepton of negative charge. Now, the question is whether helicity is the {\em only} difference between the $\nbarq (h=+1/2)$ emitted in the $W^-$ decay and the $\nu_m (h=-1/2)$ emitted in the $W^+$ decay. Would a $\nmb(h=+1/2)$ become a $\nu_m (h=-1/2)$ if we could somehow reverse its helicity? If it would, then for a given helicity,
\beq
\nmb (h) = \nu_m(h) \; .
\label{eq5}
\eeq
When we say that $\nu_m$ is identical to its antiparticle, we mean that it obeys this condition. It is then referred to as a Majorana neutrino.

It may be that the $\nmb(h=+1/2)$ emitted in $W^-$ decay and the $\nu_m (h=-1/2)$ emitted in $W^+$ decay not only have opposite helicity, but also differ from one another by a quantum number, such as the hypothetical conserved lepton number L which distinguishes a lepton from an antilepton. If $\overline{\nu_m} (h=+1/2)$ and $\nu_m (h=-1/2)$ do have such an added difference, then 
\beq
\nmb(h) \neq \nu_m(h) \; .
\label{eq6}
\eeq
That is, $\nu_m$ is not identical to its antiparticle. A neutrino which carries a conserved quantum number, so that its antiparticle will carry the opposite value of this quantum number and hence differ from the neutrino, is called a Dirac neutrino \cite{r9}.

\begin{figure}[htb]
\includegraphics[scale=0.6]{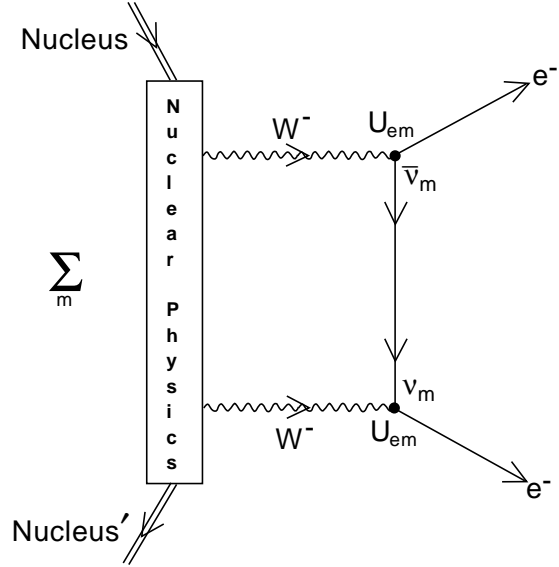}
\caption{Neutrino-exchange diagram for $\bb$. A coherent sum over the contributions of the different neutrino mass eigenstates $\nu_m$ must be performed.}
\label{f1}
\end{figure}

The most popular explanation of why neutrinos are so light, the ``see-saw mechanism'' \cite{r10}, predicts that they are Majorana particles. To try to confirm this prediction, one may look for neutrinoless double beta decay ($\bb$). This is the process Nucleus $\ra$ Nucleus$^\prime$ + 2e$^-$, in which one nucleus decays into another with the emission of two electrons and nothing else. If $\nmb (h) = \nu_m (h)$, this decay can proceed via the diagram in Fig.~\ref{f1}.
In this diagram, two virtual $W$ bosons are emitted, and then these bosons exchange one of the neutrino mass eigenstates to create the outgoing electrons. The coupling acting at the vertices where the electrons are produced is assumed to be the left-handed Standard Model (SM) weak interaction. For definiteness, we shall suppose that the exchanged neutrino is created at the upper vertex in Fig.~\ref{f1}, and absorbed at the lower vertex, as indicated by the arrows. 
Now, the SM weak interaction conserves L. Thus, if $\nu_m$ and $\nmb$ differ, then the neutral particle created at the upper vertex together with an e$^-$ must be a $\nmb$. However, when this same particle is absorbed at the lower L-conserving vertex to make the second e$^-$, it must be a $\nu_m$. Thus, if $\nmb \neq \nu_m$, this diagram cannot occur. By contrast, if $\nmb = \nu_m$, the diagram does occur. Thus, the observation of $\bb$ would demonstrate that neutrinos are Majorana particles.

The amplitude for the left-handed SM interaction to absorb the $\nu_m$ in Fig.~\ref{f1} to make an e$^-$ is small unless the $\nu_m$ helicity $h=-1/2$. However, the amplitude for this same interaction to create the $\nu_m$ at the upper vertex together with an e$^-$  but with  $h=-1/2$ is of order $M_m$\,/\,(Energy of the neutrino). Thus, in view of the factor $U_{em}$ which describes the coupling of a $\nu_m$ to an electron at each of the two weak vertices in Fig.~\ref{f1}, we conclude that, if  $\nmb = \nu_m$,
\begin{eqnarray}
\lefteqn{\mbox{Amplitude }[\bb] = }  \nonumber  \\
 & & (\sum_m M_m U^2_{em}) \times (\mbox{Nuclear Factor}) \; .
\label{eq7}
\end{eqnarray}

The factor $\sum_m M_m U^2_{em} \equiv M_{\beta\beta}$ is referred to as the effective neutrino mass for $\bb$, and the ``Nuclear Factor'' describes the very nontrivial nuclear physics of the process \cite{r11}.

To determine the elements $U_{\ell m}$ of the leptonic mixing matrix, we can appeal to experiments on neutrino oscillation. The probability $P\, (\nlbarp  \ra  \nlpbarp)$ for a neutrino (antineutrino) of energy $E$, born with flavor $\ell$, to oscillate in a distance $L$ into a neutrino (antineutrino) of flavor $\ell^\prime$ is given by 
\begin{eqnarray}
\lefteqn{P\, (\nlbarp  \ra  \nlpbarp) = }  \nonumber \\
 & & \delta_{\ell\ell^{\st \prime}} - 4 \sum_{m > m^{\st \prime}} \Re ( {\cal U}_{\ell\ell^\prime;mm^\prime}) \sin^2 (\dm2 \frac{L}{4E}) \nonumber \\
 & & \pom 2  \sum_{m > m^{\st \prime}} \Im ({\cal U}_{\ell\ell^\prime;mm^\prime}) \sin (\dm2 \frac{L}{2E}) \; ,
\label{eq8}
\end{eqnarray}
where ${\cal U}_{\ell\ell^\prime;mm^\prime} \equiv  U^*_{\ell m} U_{\ell^\prime m} U_{\ell m^\prime} U^*_{\ell^\prime m^\prime}$.
From these expressions, we see that complex phases in $U$ can lead to CP-violating differences between corresponding neutrino and antineutrino oscillation probabilities. However, suppose some oscillation involves only two neutrinos, or effectively only two neutrinos. By the latter, we mean that there are, say, three neutrino mass eigenstates $\nu_1,\; \nu_2, \; \nu_3$, but $\nu_1$ and $\nu_2$ are nearly degenerate, and act in many ways like a single neutrino in experiments whose $L/E$ is insufficiently large to reveal that $\delta M^2_{21}$ is not zero.
 One can show that when oscillation involves only two, or effectively only two, neutrinos, the oscillation probability depends only on the {\em sizes} of the matrix elements $U_{\ell m}$, and not on their phases. Thus, sizes of the $U_{\ell m}$ can be determined via suitable oscillation experiments \cite{r4a}.

The {\em phases} of combinations of $U$ elements could be determined by studying the CP-violating effects to which these phases lead. The matrix $U$ can contain two types of CP-violating phases: Universal phases, which can be present whether neutrinos are Dirac or Majorana particles, and Majorana phases, which can be present in addition to the Universal phases if neutrinos are Majorana particles. 
For a given number of neutrinos, the number of independent CP-violating phases of each type that $U$ can contain is as indicated in the following table.

\vspace{.15in}
\begin{tabular}{@{}ccc}   \hline
Number of		& Universal	&Majorana	\\
Neutrinos		& Phases		& Phases		  \\ \hline
2 & 0 & 1  \\
3 & 1 & 2  \\
4 & 3 & 3  \\  \hline
\end{tabular}
\vspace{.2in}

To understand why $U$ can contain more CP-violating phases when neutrinos are Majorana particles that it can when they are Dirac particles, let us momentarily consider {\em quark} mixing. Phases can be removed from the quark mixing matrix $V$ by redefining any quark field $q$ by multiplying it with a phase factor: $q \ra e^{i\varphi}q$. Any phase which can be removed from $V$ in this way has no physical consequences. Phases can be removed from the leptonic mixing matrix in a similar fashion. However, when $\nmb = \nu_m$, then, roughly speaking, the $\nu_m$ field is its own charge conjugate. That is, 
\beq
\mbox{Charge conjugate } (\nu_m) \equiv C \,\nmb^T = \nu_m \; ,
\label{eq9}
\eeq
where $C$ is the charge conjugation matrix and $T$ stands for transpose. Obviously, we are not free to subject a $\nu_m$ field which must satisfy \Eq{9} to the redefinition $\nu_m \ra e^{i\varphi} \nu_m$, because under this redefinition $\nmb \ra e^{-i\varphi} \nmb$, and \Eq{9} is no longer satisfied. Thus, some of the phases which can be removed from $U$ when neutrinos are Dirac particles like quarks cannot be removed when neutrinos are Majorana particles. 
These unremovable ``Majorana'' phases have physical CP-violating consequences. However, it can be shown that the Majorana CP-violating phases affect $\bb$, but not neutrino oscillation. In contrast, the Universal CP-violating phases affect neutrino oscillation, but not $\bb$. Thus, one seeks evidence of these two kinds of CP-violating phases in two quite different kinds of experiments.

In neutrino oscillation, the CP-violating observable is the difference
\beq\DCP{\ell\ell^{\st \prime}} \equiv P (\nu_\ell \ra \nu_{\ell^{\sss \prime}}) - P (\overline{\nu_\ell} \ra \overline{\nu_{\ell^{\sss \prime}}}) \; ,
\label{eq10}
\eeq
where $\ell^{\st \prime} \neq \ell$. Interestingly enough, if there are only three neutrinos, then 
\begin{eqnarray}
\lefteqn{\DCP{e\mu} = \DCP{\mu\tau} = \DCP{\tau e}}  \nonumber \\
 & & = 16 Js_{12}s_{23}s_{31} \; ,
\label{eq11}
\end{eqnarray}
where $J \equiv \Im (U_{e1} U^*_{e2} U^*_{\mu 1} U_{\mu 2})$, and $s_{mm^{{\st \prime}}} \equiv \sin (\dm2 L/4E)$. That is, the CP-violating differences $\Delta_{CP}$ in the three different flavor oscillations that one can study are all equal, and they all depend on the CP-violating parameter $J$, which obviously is a measure of the degree to which the elements $U_{\ell m}$ are not real. 
To observe the differences $\Delta_{CP}$, one must clearly do experiments with $L/E$ large enough that $\dm2 L/4E$ is not too tiny, even for the smallest of the splittings $\dm2$. These experiments would be both challenging and very interesting \cite{r12}.

As we have discussed, the observation of $\bb$ would establish that neutrinos are Majorana particles. In addition, a measured rate for this process, combined with a calculated value of the corresponding Nuclear Factor, would determine $\Mbb \equiv |\sum_m M_m U^2_{em}|$ via \Eq{7}. $M_{\beta\beta}$ is a different combination of neutrino masses than the (Mass)$^2$ splittings measured in neutrino oscillation. Thus the value of $\Mbb$ could test the neutrino mass spectra suggested by the oscillation data in ways that the latter data cannot \cite{r13}. 
\begin{figure}[b]
\includegraphics[scale=0.6]{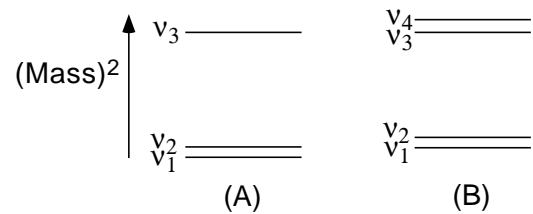}
\caption{Neutrino mass spectra suggested by neutrino oscillation data.}
\label{f2}
\end{figure}
For example, a value $\Mbb \gsim 0.03$ eV would exclude a 3-neutrino mass hierarchy like that pictured in Fig.~\ref{f2}(A), where it is assumed that $M_3$, the mass of the heaviest neutrino $\nu_3$, is no larger than necessary to explain the (Mass)$^2$ splittings implied by the oscillation data. 
$\Mbb \gsim 0.03$ eV would also exclude a 4-neutrino scheme of the type in Fig.~\ref{f2}(B), where it is assumed that $M_3$ and $M_4$, the masses of the heavy neutrinos $\nu_3$ and $\nu_4$, are no larger than necessary to explain the (Mass)$^2$ splittings implied by the oscillation data, and it is also assumed that the light neutrinos $\nu_1, \nu_2$, but not the heavy ones, couple appreciably to the electron. A value $\Mbb \gsim 0.03$ eV would allow a 4-neutrino spectrum like that of Fig.~\ref{f2}(B) but where the heavy neutrinos $\nu_3, \nu_4$ are the ones that couple appreciably to the electron. For such a spectrum,
\beq
\Mbb = \sqrt{\delta M^2_{\mbox{\ss LSND}}} \sqrt{1 - \sin^2 2\theta_\odot \sin^2 \alpha_{CP}} \; ,
\label{eq12}
\eeq
where $\delta M^2_{\mbox{\ss LSND}}$ is the (Mass)$^2$ splitting called for by the LSND oscillation signal, $\theta_\odot$ is the mixing angle governing solar neutrino oscillation, and $\alpha_{CP}$ is one of the extra Majorana CP-violating phases that can occur in the $U$ matrix when neutrinos are Majorana particles. Once $\delta M^2_{\mbox{\ss LSND}}$ and $\theta_\odot$ are known, a measurement of $\Mbb$ and the application of \Eq{12} would determine $\alpha_{CP}$.

Assuming CPT invariance, the magnetic or electric dipole moment of a neutrino $\nu_m$, and that of its antiparticle $\nmb$, must be equal and opposite. Thus, if $\nu_m$ is a Majorana neutrino, so that $\nmb = \nu_m$, then its magnetic and electric dipole moments must vanish. Only a Dirac neutrino can have {\em non-transition} dipole moments. However, both Dirac and Majorana neutrinos can have {\em transition} dipole moments, which lead to the transitions $\nu_m \ra \nu_{m^\prime \neq m} + \gamma$.

In simple extensions of the Standard Model, the magnetic dipole moment $\mu_m$ of a Dirac neutrino $\nu_m$ is very small because it is proportional to the mass of $\nu_m$, $M_m$. This proportionality is due to the fact that the coupling of a neutrino dipole moment to a photon must flip the neutrino handedness, and the only ingredient in the Standard Model diagram for this coupling which can cause this flip is the neutrino mass \cite{r14}. The proportionality constant is not large; indeed, one finds that \cite{r15}
\beq
\mu_m = 3.2\times 10^{-19} M_m\mbox{(eV )}\mu_{\mbox{{\ss Bohr}}} \; ,
\label{eq13}
\eeq
where $\mu_{\mbox{{\ss Bohr}}}$ is the Bohr magneton. Similar considerations limit the size of any electric dipole moment. The latter cannot arise without a violation of CP, and consequently is suppressed even further by this circumstance.

To be sure, models which permit much bigger dipole moments have been constructed. One can look for neutrino dipole moments by seeking a dipole-moment contribution to the scattering of either reactor or solar \cite{r16} neutrinos from electrons.

The lifetimes of neutrinos may well be determined by their radiative decays, $\nu_m \ra \nu_{m^\prime} + \gamma$, which are their simplest decays into known particles. If so, then these lifetimes are extremely long compared, say, to one second. The reason is that the radiative decays cannot occur at tree level, and, like the neutrino dipole moments, they are suppressed by the smallness of neutrino masses. Exotic decay modes may yield shorter lifetimes \cite{r17}. If there should be a mass eigenstate $\nu_m$ which is a component of $\nu_\mu$ and which decays in less than a second, then the atmospheric neutrino data, which are usually explained in terms of $\nu_\mu \ra \nu_\tau$ oscillation, can be equally well explained in terms of the decay of this $\nu_m$ \cite{r18}. Of course, even with this alternative explanation, the atmospheric data still imply that neutrinos have mass.

While the evidence for neutrino mass has become quite convincing, we are just beginning to learn how many neutrinos there are, how much they weigh, their nature, and their mixings. We will learn a lot about these matters from experimental results expected to appear in the coming years---years which will be a very exciting time in neutrino physics.

\vspace{.2in}  %\hspace{-.25\baselineskip}
{\bf Acknowledgments}  \newline
It is a pleasure to thank the Fermilab Theoretical Physics Group for its warm hospitality while a part of this talk was written.


\begin{thebibliography}{99}
\bibitem{r11} K. Balaji and W. Grimus, in \prdn{62}{00}{053009}, have recently proposed that we might be able to tell whether neutrinos are of Majorana or Dirac character by comparing flavor survival probabilities of left- and right-handed neutrinos traveling in a constant magnetic field. This is an interesting idea, but it remains to be seen whether this approach will prove practical.
\end{thebibliography}
\end{document}